\documentclass[english,aps,prl,twocolumn,superscriptaddress]{revtex4-1}

\usepackage{bm}
\usepackage{appendix}
\usepackage{graphicx}
\usepackage{amsmath}
\usepackage{amssymb}%
\usepackage{color}
\usepackage{lipsum}
\usepackage{ulem}
\usepackage[colorlinks=true,citecolor=blue,urlcolor=blue]{hyperref}
\def\cp#1{\mathbf{#1}}

\begin{document}
\title{Non-Bloch quench dynamics}

\author{Tianyu Li}
\affiliation{CAS Key Laboratory of Quantum Information, University of Science and Technology of China, Hefei 230026, China}
\author{Jia-Zheng Sun}
\affiliation{CAS Key Laboratory of Quantum Information, University of Science and Technology of China, Hefei 230026, China}
\author{Yong-Sheng Zhang}
\email{yshzhang@ustc.edu.cn}
\affiliation{CAS Key Laboratory of Quantum Information, University of Science and Technology of China, Hefei 230026, China}
\affiliation{CAS Center For Excellence in Quantum Information and Quantum Physics, Hefei 230026, China}
\author{Wei Yi}
\email{wyiz@ustc.edu.cn}
\affiliation{CAS Key Laboratory of Quantum Information, University of Science and Technology of China, Hefei 230026, China}
\affiliation{CAS Center For Excellence in Quantum Information and Quantum Physics, Hefei 230026, China}

\begin{abstract}
We study the quench dynamics of non-Hermitian topological models with non-Hermitian skin effects. Adopting the non-Bloch band theory and projecting quench dynamics onto the generalized Brillouin zone, we find that emergent topological structures, in the form of dynamic skyrmions, exist in the generalized momentum-time domain, and are correlated with the non-Bloch topological invariants of the static Hamiltonians.
The skyrmion structures anchor on the fixed points of dynamics whose existence are conditional on the coincidence of generalized Brillouin zones of the pre- and post-quench Hamiltonians. Global signatures of dynamic skyrmions, however, persist well beyond such a condition, thus offering a general dynamic detection scheme for non-Bloch topology in the presence of non-Hermitian skin effects.
Applying our theory to an experimentally relevant, non-unitary quantum walk, we explicitly demonstrate how the non-Bloch topological invariants can be revealed through the non-Bloch quench dynamics.
\end{abstract}

\maketitle

Non-Hermitian Hamiltonians arise in open systems~\cite{QJ,benderreview}, and have attracted significant attention in recent years~\cite{bender,photonpt1,review1,review2,review3,nonHtopo1,nonHtopo2,nonHtopo3}.
As a peculiar feature of a wide class of non-Hermitian Hamiltonians, their nominal bulk eigenstates are localized near boundaries under what is now known as the non-Hermitian skin effects~\cite{WZ1,WZ2,Budich,mcdonald,alvarez,murakami,ThomalePRB,fangchenskin,kawabataskin,Slager,XZ,XR}. Remarkably, for non-Hermitian topological models with skin effects, the conventional bulk-boundary correspondence breaks down~\cite{Lee,WZ1,WZ2,Budich,Slager,murakami,ThomalePRB,XZ,XR}, whose restoration calls for a non-Bloch band theory where
topological invariants are calculated in the generalized Brillouin zone (GBZ) under relevant boundary conditions~\cite{WZ1,WZ2,murakami,tianshu,tianyu,XZ,ThomalePRB,fangchenskin}, rather than the conventional Brillouin zone (BZ) under a periodic boundary condition (PBC). Both non-Hermitian skin effects and non-Bloch bulk-boundary correspondence have recently been experimentally confirmed~\cite{teskin,teskin2d,metaskin,photonskin,scienceskin,EP}. However, a direct detection of non-Bloch topological invariants is yet to be demonstrated.
Here we show how non-Bloch topological invariants can be determined through dynamic topological structures in quantum quenches that are experimentally accessible.

A quantum quench describes the evolution of an eigenstate of the initial Hamiltonian $H^i$ driven by a final Hamiltonian $H^f$. Particularly, for one-dimensional topological systems, previous studies have shown that dynamic skyrmions, a topological structure hinged upon fixed points of dynamics, should appear in the emergent momentum-time domain, when the initial and final Hamiltonians possess distinct topological properties~\cite{Chen17,Ueda17,iS}. It follows that by studying the dynamic signatures of a quantum quench, one gains information regarding the topological invariants of the pre- and post-quench Hamiltonians~\cite{Gogolin15, Bhaseen15, Dora15, BH16, Balatsky, Budich16, Refael16, Bhaseen16, Zhai17, Xiongjun,Heyl, longwen,jinlong1,jinlong2, Weitenberg17,chenshuai,Weitenberg18,XPdqpt,XPNC}.
Similar conclusions hold for non-unitary quenches governed by non-Hermitian Hamiltonians, but are predicated upon two conditions~\cite{iS}:
i) the decoupling of dynamics in different momentum sectors under the lattice translational symmetry; and ii) reality of the eigenenergy spectra of both $H^i$ and $H^f$, where the parity-time (PT) symmetry plays an important role~\cite{bender,photonpt1,benderreview,review1,review2}.
For a non-Hermitian topological model with skin effects, however, a compromise between these two requirements seems to be a tall order. On one hand, dynamics of different momentum states are no longer decoupled under an open boundary condition (OBC). On the other, for a system possessing non-Hermitian skin effects, the associated Bloch spectra in the conventional BZ necessarily form loops in the complex plane~\cite{fangchenskin,kawabataskin}, such that real eigenenergy spectra only exist in the GBZ under OBCs, protected by a non-Bloch PT symmetry~\cite{nonBPT1,nonBPT2,EP}.

We circumvent these issues by projecting the quench dynamics onto the generalized momentum sectors of the GBZ. Such a non-Bloch analysis enables us to reveal, in the generalized momentum-time domain, dynamic skyrmions that are intimately related to the non-Bloch topological invariants of the pre- and post-quench Hamiltonians. In particular, for a non-Hermitian Su-Schrieffer-Heeger (SSH) model with asymmetric hopping and under OBC, we prove that the existence of the underpinning fixed points for dynamic skyrmions exist, provided $H^i$ and $H^f$ are both in the non-Bloch PT unbroken phase (i.e., both having real eigen spectra) and feature the same GBZ, but with distinct non-Bloch winding numbers. When the GBZs of $H^i$ and $H^f$ are different, fixed points only exist in a perturbative sense, whereas the global signatures of dynamic skyrmions persist, allowing for the detection of non-Bloch winding numbers over a wide parameter regime.
We apply our theory to the recently implemented, non-unitary topological quantum walk, and illustrate the extraction of non-Bloch winding number from the non-Bloch quench dynamics.


{\it Quenching in the GBZ:---}
We first consider the quench dynamics of a non-Hermitian SSH model~\cite{SSH,WZ1}
\begin{align}
H&=\sum_{n}\Big[(t_{1}+\gamma)|n,A\rangle\langle n,B|+(t_{1}-\gamma)|n,B\rangle\langle n,A|\nonumber\\
&+t_{2}|n,B\rangle\langle n+1,A|+t_{2} |n+1,A\rangle\langle n,B|\Big],\label{eq:SSH}
\end{align}
where each unit cell (labeled $n$) has two sublattice sites (labeled $A$ and $B$), with intra- and inter-cell hopping rates given by $t_1\pm\gamma$ and $t_2$, respectively. The asymmetric intra-cell hopping here gives rise to non-Hermitian skin effects, such that
topological properties of the model under OBC are characterized by the non-Bloch band theory, where a GBZ must be invoked, due to the deviation of the nominal bulk states from extended Bloch waves.

Specifically, the GBZ of Hamiltonian (\ref{eq:SSH}) under OBC is a circle with radius $r=\sqrt{|(t_1-\gamma)/(t_1+\gamma)|}$ on the complex plane~\cite{WZ1,supp}, where we denote the azimuth angle as $k\in (\pi,\pi]$. In the Hermitian limit with $\gamma=0$, the GBZ reduces to the conventional BZ, a unit circle on the complex plane with $r=1$, with $k$ being the Bloch quasi-momentum.
As the non-Hermitian, non-Bloch counterparts of the Bloch states, we then introduce the biorthogonal spatial basis in a generalized momentum space characterized by $k$: $|\beta_{k,R(L)} \rangle=\frac{1}{\sqrt{N}}\sum_{n} r^{\pm n} e^{ikn}|n\rangle$ ($N$ is the total number of unit cells, $\alpha\in \{A,B\}$), which satisfy the biorthogonal and completeness relations~\cite{DCB}: $\langle\beta_{k,L}|\beta_{k',R} \rangle=\delta_{k,k'}$ and $\sum_{k}|\beta_{k,R}\rangle \langle\beta_{k,L}|=\cp{1}_n$, with $\cp{1}_n=\sum_n|n\rangle\langle n|$. Conveniently, while $|\beta_{k,R/L}\rangle$ become extended Bloch states in the Hermitian limit, $|\beta_{k,R}\rangle$ ($|\beta_{k,L}\rangle$) is localized toward the right (left) boundary for $r>1$, and vice versa, thus constituting a natural basis for both Hermitian and non-Hermitian cases.

To correctly characterize the non-Bloch topology, we define the projection operator $P_k=|\beta_{k,R}\rangle\langle \beta_{k,L}|\otimes \cp{1}_\alpha$, $\cp{1}_\alpha$ being the identity operator in the sublattice basis $\{|A\rangle, |B\rangle\}$, and project
Hamiltonian (\ref{eq:SSH}) onto the GBZ as $H_k=P_k H P_k$. Writing $H_k=\sum_i d_i(k)\sigma_i$ ($\sigma_i$ are the Pauli operators with $i=1,2$), the non-Bloch winding number is
\begin{align}
\nu=\frac{1}{2\pi}\int dk \frac{-d_2\partial_k d_1+d_1\partial_k d_2}{d_1^2+d_2^2},\label{eq:nonblochnu}
\end{align}
which recovers the bulk-boundary correspondence under OBC~\cite{WZ1}. Note that while the eigen spectrum under PBC is always complex, it is completely real under OBC for $|t_1|>\gamma$, due to the presence of a non-Bloch PT symmetry~\cite{nonBPT1,nonBPT2}. Following Eq.~(\ref{eq:nonblochnu}), the non-Bloch topological phase boundaries in this PT-unbroken region are given by $t_2=\pm \sqrt{t_1^2-\gamma^2}$, which reduce to those of a Hermitian SSH model when $\gamma=0$, with Eq.~(\ref{eq:nonblochnu}) then yielding the Bloch winding number.

\begin{figure}[tbp]
\centering
\includegraphics[width=3.2in]{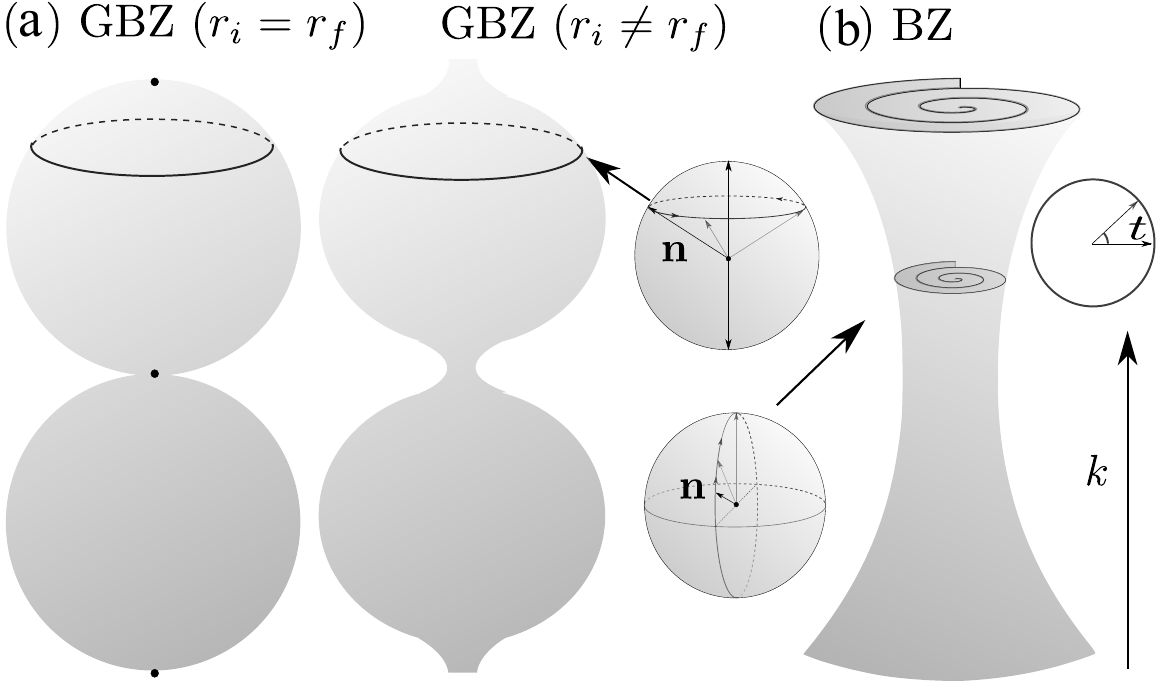}
\caption{Schematic illustration of (a) non-Bloch and (b) Bloch quench dynamics of the non-Hermitian SSH model of Eq.~(\ref{eq:SSH}). (a) Non-Bloch quench dynamics in the GBZ is periodic in each $k$-sector, with either (left) identical GBZs or (right) distinct GBZs for the pre- and post-quench Hamiltonians~\cite{Chen17,iS}. The radius $r$ and generalized momentum $k$ of the GBZ, as well as the Bloch-sphere vector $\bm{n}(k,t)$ are defined in the main text. Black dots in the left panel indicate fixed points. (b) Bloch quench dynamics is steady-state approaching in the $k$-sectors.}
\label{fig:fig1}
\end{figure}

In a general quantum quench, an eigenstate of the initial Hamiltonian $H^i$ evolves under the final Hamiltonian $H^f$.
For an initial state $|\psi^i_k\rangle$ in a given $k$-sector, for instance,
the time-dependent density matrix $\rho(t)$ evolves according to
\begin{align}
\rho(t)=e^{-iH^f t}|\Psi^i_k\rangle\langle\Psi^i_k| e^{iH^{f\dag} t},\label{eq:rhoevolve}
\end{align}
where $|\Psi^i_k\rangle=|\beta^i_{k,R}\rangle\otimes |\psi^i_{k}\rangle$, with $|\beta^i_{k,R}\rangle$ the right spatial basis of $H^i$ and
$|\psi^i_{k}\rangle$ the internal state in the sublattice basis.
Unlike the quasi-momentum in the BZ, generalized momentum $k$ in a GBZ is not a good quantum number,
as $\langle \beta^f_{k,L}|H^f|\beta^i_{k',R}\rangle\neq 0$ for $\gamma\neq 0$, such that a state initialized in a $k$-sector inevitably proliferates into other $k$-sectors in the time evolution.
However, it can be shown that for $r_i=r_f$, i.e., the GBZs of $H^i$ and $H^f$ coincide with each other, these cross-coupling terms vanish in the thermodynamic limit, rendering the dynamics block-diagonal in the GBZ, similar to the conventional quench dynamics in the BZ.

\begin{figure*}[tbp]
\centering
\includegraphics[width=7in]{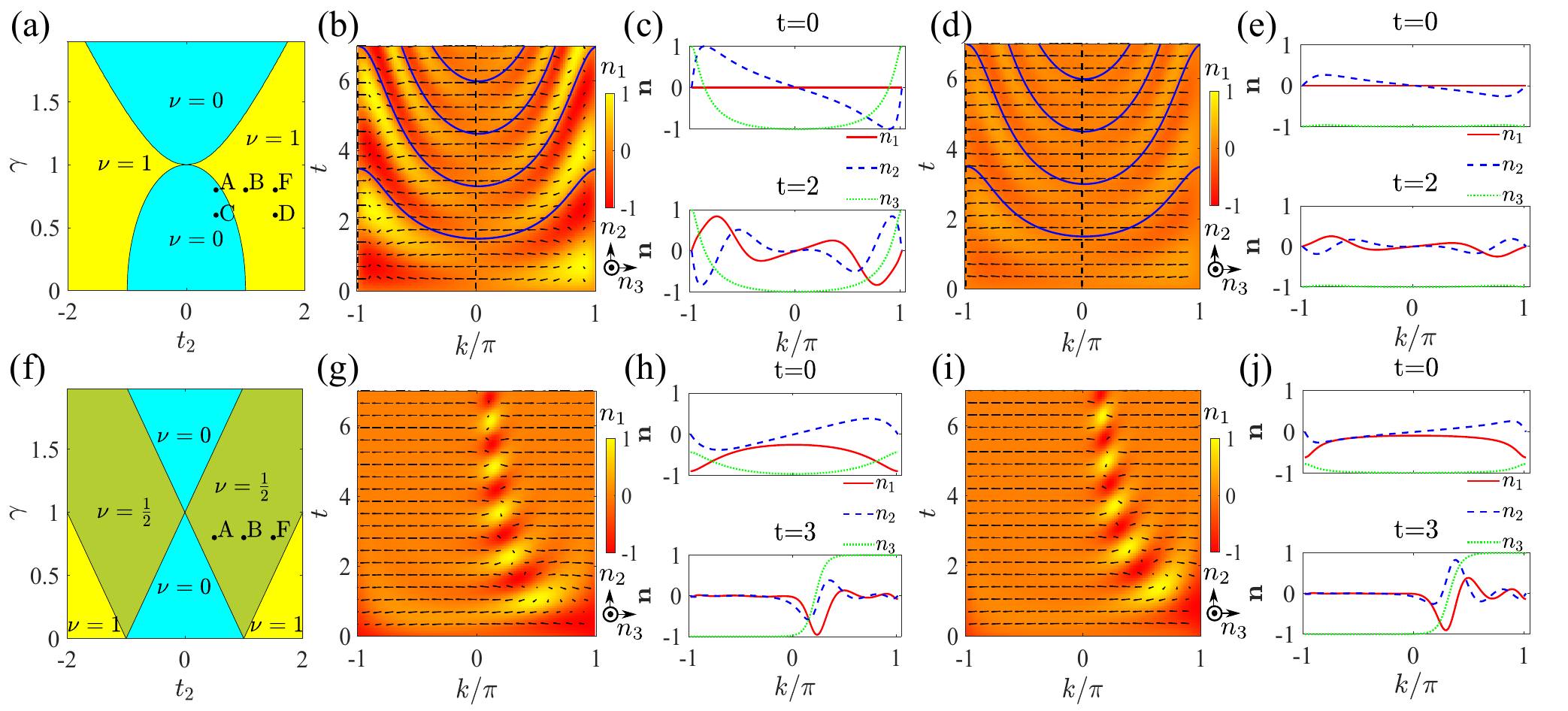}
\caption{Quench dynamics projected in (top row) the GBZ and (lower row) BZ, respectively, where $t_1$ is taken as the unit of energy. (a) Non-Bloch topological phase diagram under OBC. All quenches are performed in the region with $\gamma<1$, where the eigen spectra are completely real.
(b) Dynamic spin texture $\bm{n}(k,t)$ in the generalized momentum-time space, for a quantum quench from point A ($H^i$ with non-Bloch winding number $\nu_i=0$) to F ($H^f$ with non-Bloch winding number $\nu_f=1$). (c) $\bm{n}(k,t)$ across the GBZ at different times $t$ for the quench in (b). (d) $\bm{n}(k,t)$ for a quench from point B ($H^i$ with non-Bloch winding number $\nu_i=1$) to F. No skyrmion structures are present, in contrast to (b). (e) $\bm{n}(k,t)$ across the GBZ at different times $t$ for the quench in (d). (f) Bloch topological phase diagram under PBC. Eigen spectra are complex throughout the phase diagram. (g) $\bm{n}(k,t)$ in the momentum-time space, for a quantum quench from point A ($H^i$ with Bloch winding number $\nu_i=1/2$) to F ($H^f$ with Bloch winding number $\nu_f=1/2$). (h) $\bm{n}(k,t)$ across the BZ at different times $t$ for the quench in (g).
(i) $\bm{n}(k,t)$ in the momentum-time space for a quench from point B to F. (j) $\bm{n}(k,t)$ across the BZ at different times $t$ for the quench in (i). Dashed vertical lines in (b)(d) denote the locations of fixed points at $k_m=0,\pm\pi$.
The parameters for points A, B, and F are $t_2=0.5,1,1.5$, respectively, with $\gamma=0.8$. Points C and D are used in Fig.~\ref{fig:fig3}, with parameters $t_2=0.5,1.5$, respectively, and $\gamma=0.6$ for both. For the non-Bloch quench processes (top), $r_i=r_f=0.33$; whereas for the Bloch quenches (lower), $r_i=r_f=1$.
}
\label{fig:fig2}
\end{figure*}

Inspired by such an observation, we first focus on the simpler case of $r_i=r_f$, and discuss more general scenarios of $r_i\neq r_f$ or non-circular GBZs later. To characterize quench dynamics in the GBZ, we first project $\rho(t)$ onto the GBZ of $H^f$~\cite{supp}
\begin{align}
\rho(k,t)=P_k\rho(t)P_k^\dag,\label{eq:proj}
\end{align}
where $P^\dag_k$ is introduced to accommodate the time evolution of $\langle \Psi_k^i|$ in Eq.~(\ref{eq:rhoevolve}). We characterize dynamics in each $k$-sector with the dynamic spin texture $\bm{n}(k,t)$ in a generalized momentum-time space spanned by $(k,t)$~\cite{XPNC,iS}
\begin{align}
\bm{n}(k,t)=\frac{\text{Tr}[\rho(k,t)\eta\bm{\tau}]}{\text{Tr}[\rho(k,t)\eta]},\label{eq:nvec}
\end{align}
where $\bm{\tau} =(\tau_1, \tau_2, \tau_3)$, with $\tau_i=\sum_{\mu\nu=\pm}|\psi^f_{k,\mu}\rangle \sigma^{\mu\nu}_i \langle \chi^f_{k,\nu}|$ (i=1, 2, 3). Here, $\sigma^{\mu\nu}_i$ are the elements of Pauli matrices $\sigma_i$, and $\sigma_0$ is a $2\times 2$ identity matrix.
$|\psi^f_{k,\pm}\rangle$ ($\langle \chi^f_{k,\pm}|$) is the right (left) eigenstate of $H^f_k$, with
$H^f_k|\psi^f_{k,\pm}\rangle=E^f_{k,\pm}|\psi^f_{k,\pm}\rangle$ ($H^{f\dag}_k|\chi^f_{k,\pm}\rangle=E^{f\ast}_{k,\pm}|\chi^f_{k,\pm}\rangle$), and $\pm$ are the band indices.
The metric operator $\eta=\sum_{\mu}|\chi^f_{k,\mu}\rangle\langle \chi^f_{k,\mu}|$ is introduced to normalize $\rho(k,t)$, such that $\bm{n}(k,t)$ is a unit, real vector that can be visualized on the Bloch sphere $S^2$~\cite{DCB,iS}.

Without loss of generality, we consider the initial state in each $k$-sector to lie within the lower band $|\psi^i_{k,-}\rangle$, with $H^i_k|\psi^i_{k,-}\rangle=E^i_{k,-}|\psi^i_{k,-}\rangle$. Importantly, when $E^f_{k,\mu}$ is real, $\bm{n}(k,t)$ rotates around the north pole of the Bloch sphere with a period $T_k=\pi/E^f_k$ [see Fig.~\ref{fig:fig1}(a) for schematics]~\cite{iS,supp}.
Consequently, dynamic fixed points $k_m$ in the GBZ can occur when $c_{\pm}=\langle \chi^f_{k_m,\pm}|\psi^i_{k_m,-}\rangle=0$, i.e., when the initial $\bm{n}(k_m,t=0)$ vector is aligned with the north ($c_-=0$) or south ($c_+=0$) pole of the Bloch sphere, leading to stationary evolutions at $k_m$.

For the quench dynamics of a non-Hermitian SSH model without skin effects, it has been shown that the presence of fixed points are related to the Bloch winding numbers of the pre- and post-quench Hamiltonians, provided both Hamiltonians belong to the PT-unbroken regime with completely real eigen spectra~\cite{XPNC,iS}. Here, it is straightforward to show that, when $r_i=r_f$, the same conclusions still hold, albeit now the fixed points are only visible in the GBZ, and the protecting symmetry for real eigen spectra is instead the non-Bloch PT symmetry. By contrast, if the quench dynamics is projected onto the BZ, i.e., replacing spatial basis states $|\beta^f_{k,R/L}\rangle$ in the projector $P_k$ with Bloch states, the dynamics is always steady-state approaching, as Bloch spectra $E^f_{k,\pm}$ are necessarily complex for $k\in \text{BZ}$ with $r^f=1$. As illustrated in Fig.~\ref{fig:fig1}(b), $\bm{n}(k,t)$ then asymptotically approaches the north ($\text{Im} E^f_k>0$) or south ($\text{Im} E^f_k<0$) pole on the Bloch sphere~\cite{supp}.

In Fig.~\ref{fig:fig2}(b)(c), we show $\bm{n}(k,t)$, in the generalized momentum-time domain, for a quench between Hamiltonians with different non-Bloch winding numbers.
While the fixed points ($k_m=0,\pm\pi$) and periodic dynamics (blue curves indicating multiples of $T_k$) divide the generalized momentum-time space into various submanifolds, arrays of momentum-time skyrmions can be identified, which are characterized by dynamic Chern numbers in the corresponding submanifolds~\cite{Chen17, Ueda17,supp}.
Such a dynamic topological structure, however, is absent when one performs the quench in the BZ under the same parameters [see Fig.~\ref{fig:fig2}(g)(h)], where the spin dynamics is essentially steady-state approaching.
For comparison, we show in Fig.~\ref{fig:fig2}(d)(e) and Fig.~\ref{fig:fig2}(i)(j), quench dynamics in the GBZ and BZ, respectively, between Hamiltonians with the same winding numbers. While fixed points still exist in the GBZ, the dynamic Chern numbers vanish on all submanifolds in
Fig.~\ref{fig:fig2}(d).
Finally, it is worth noting that, in our analysis, projected quench dynamics in different $k$-sectors are calculated independently. For a half-filled system initialized in the lower band, however, additional cross-coupling terms from other $k$-sectors would appear, but which are typically negligibly small under all parameters considered here~\cite{supp}.

\begin{figure}[tbp]
\centering
\includegraphics[width=3.2in]{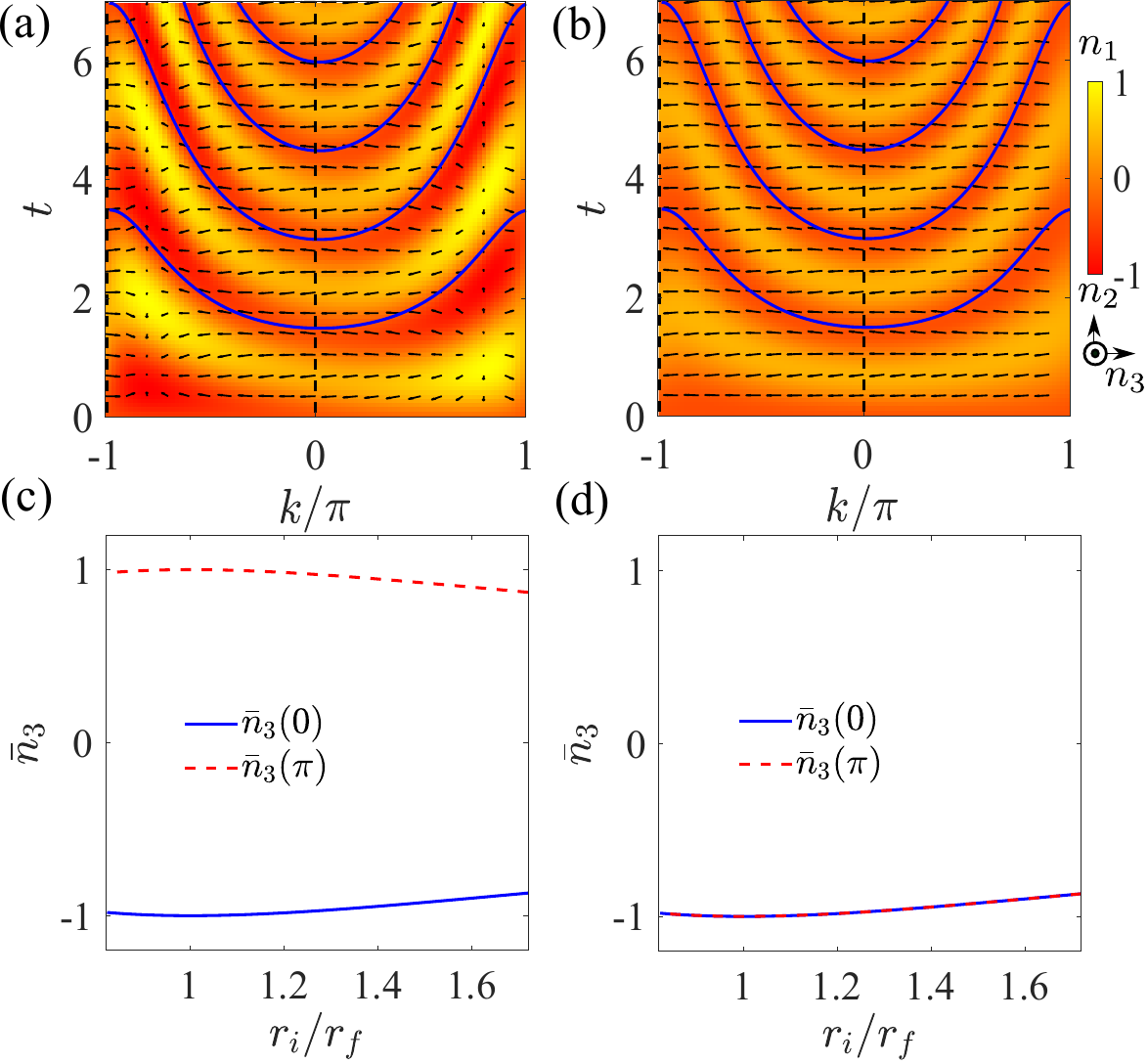}
\caption{Non-Bloch quench for $r_i\neq r_f$. (a) Spin texture $\bm{n}(k,t)$ in the generalized momentum-time space, for a quench from point C in Fig.~\ref{fig:fig2}(a) ($\nu_i=0$, $r_i=0.5$) to point F ($\nu_f=1$, $r_f=0.33$).
(b) Spin texture $\bm{n}(k,t)$ in the generalized momentum-time space, for a quench from point D in Fig.~\ref{fig:fig2}(a) ($\nu_i=1$, $r_i=1$) to point F.
(c) Time-averaged spin components $\bar{n}_3(k=0)$ and $\bar{n}_3(k=0)$ as functions of varying $r_i/r_f$, for quench processes ending at F and starting from points close to C, with $\nu_i=0$ and a fixed $t_2=0.5$. (d) $\bar{n}_3(0)$ and $\bar{n}_3(\pi)$ as functions of varying $r_i/r_f$, for quench processes ending at F and starting from points close to D, with $\nu_i=1$ and a fixed $t_2=1.5$. We take $t_1=1$ as the unit of energy.
}
\label{fig:fig3}
\end{figure}

{\it Beyond the simple case:---}
While the GBZ of Hamiltonian (\ref{eq:SSH}) under OBC is always circular in the complex plane, it is easy to choose parameters such that $r_i\neq r_f$. In this case, dynamics in different $k$-sectors of the GBZ are coupled. Nevertheless, one can still perform the projection in Eq.~(\ref{eq:proj}), and focus solely on dynamics in different $k$-sectors of $H^f$. While fixed points no longer rigorously exist in this case [see Fig.~\ref{fig:fig1}(a) for schematics]~\cite{supp}, remnants of the dynamic skyrmions for $r_i/r_f=1$ (for $\nu_i\neq\nu_f$), or the lack thereof (for $\nu_i=\nu_f$), would carry over and leave identifiable signatures in the dynamic spin texture, over a considerable range of $r_i/r_f$.

We numerically confirm such a picture in Fig.~\ref{fig:fig3}(a)(b), where $r_i/r_f=1.5$. For both cases, the system is initialized in the lower-band with occupied $|\psi^i_{k,-}\rangle$ of each $k$-sector, and under the same parameters with $\nu_i=0$. We then quench the system under final Hamiltonians featuring $\nu_f=1$ and $\nu_f=0$, respectively. As expected, skyrmion-like structures only emerge in Fig.~\ref{fig:fig3}(a) where $\nu_i\neq \nu_f$. To capture key features of these structures away from $r_i/r_f=1$, we define the time-averaged spin component $\bar{n}_3(k)=\frac{1}{T_k}\int_0^{T_k} n_3(k,t) dt$, which should give $\bar{n}_3(\pi)=-\bar{n}_3(0)\approx 1$ in the presence of skyrmion-like structures, and $\bar{n}_3(0)=\bar{n}_3(\pi)\approx 0$ otherwise.
In Fig.~\ref{fig:fig3}(c)(d), we show $\bar{n}_3(k)$ at $k=0$ and $k=\pi$ over a range of different $H^i$ (hence varying $r_i/r_f$), where
the presence [see Fig.~\ref{fig:fig3}(c)] and absence [see Fig.~\ref{fig:fig3}(d)] of skyrmion-like structures can be clearly identified.
Crucially, knowing $\bar{n}_3(k_m)$ and one of the non-Bloch winding numbers $\nu^{i/f}$, one can immediately infer the other.
Thus, while dynamic Chern numbers cannot be rigorously defined for $r_i/r_f\neq 1$, the remnant of dynamic skyrmions, due to their robustness as a global topological structure in the generalized momentum-time space, can still be used to determine non-Bloch topological invariants over a considerable range of $r_i/r_f$. Remarkably, similar conclusions can be drawn for quench dynamics between Hamiltonians with distinct, non-circular GBZs~\cite{supp}.

\begin{figure}[tbp]
\centering
\includegraphics[width=3.2in]{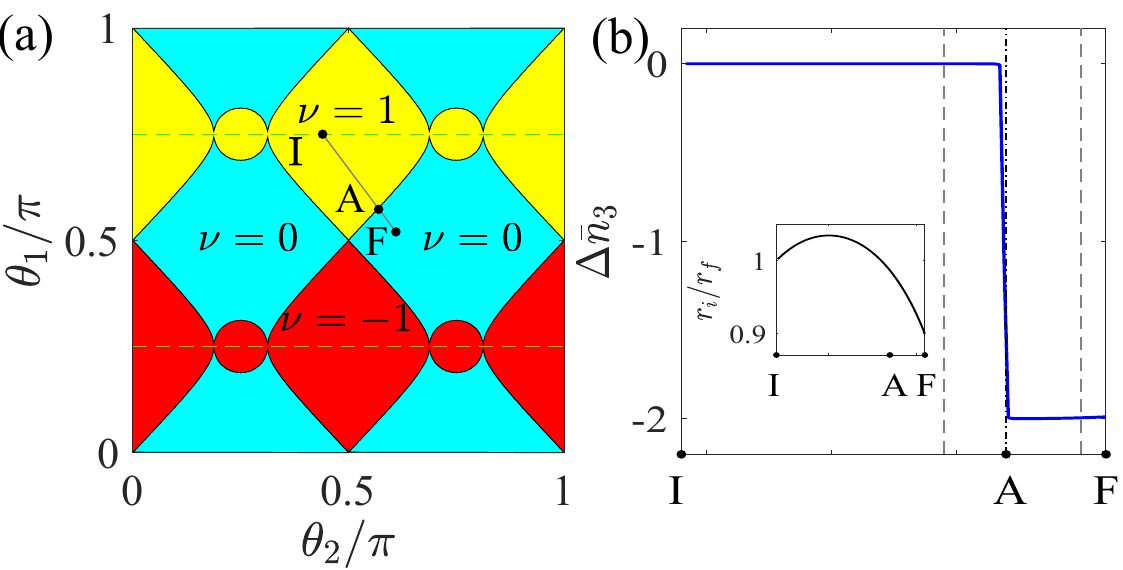}
\caption{(a) Non-Bloch topological phase diagram under OBC for the effective Hamiltonian $H$ of the Floquet operator $U$ in Eq.~(\ref{eq:U}). The horizontal green dashed lines are locations where quasi-local eigenstates exist. (b) $\Delta\bar{n}_3$ (see main text for definition) evaluated for quench processes starting from $I$, and ending at different points along the line I--F in (a). The vertical dashed and dash-dotted lines indicate, respectively, the Bloch and non-Bloch topological phase boundaries.
Inset: variation of $r_i/r_f$ as $H^f$ changes from I to F.
The parameters for points I, A and F are $(\theta_1,\theta_2)=(0.75\pi,0.44\pi), (0.573\pi, 0.57\pi), (0.52\pi, 0.61\pi)$, respectively.
}
\label{fig:fig4}
\end{figure}

{\it Non-unitary quantum walk:---}
Quantum-walk dynamics is an ideal platform for demonstrating non-Bloch quenches discussed here, particularly in light of the recent experimental observation of dynamic skyrmion structures therein~\cite{XPNC}. For such a purpose, we consider a non-unitary quantum walk along a one-dimensional lattice, governed by the Floquet operator~\cite{photonskin}
\begin{equation}
U=R(\theta_{1})S_{2}R(\theta_{2})MR(\theta_{2})S_{1}R(\theta_{1}),\label{eq:U}
\end{equation}
where the coin operator
$R(\theta_{i})={\bf{1}}_{w}\otimes e^{-i\theta_{i}\sigma_{y}}$ ($i=1, 2$),
and the gain-loss operator
$M={\bf{1}}_{w} \otimes (e^{\gamma} |A\rangle \langle A|+e^{-\gamma} |B\rangle \langle B|)$. Here ${\bf{1}}_{w}=\sum_{x}|x\rangle \langle x| $, $x$ labels the lattice sites, $\{|A\rangle,|B\rangle\}$ are the internal coin states, and $\gamma$ is the gain-loss parameter.
The shift operator
$S_{1}=\sum_{x}|x\rangle \langle x| \otimes |A\rangle \langle A|+ |x+1\rangle \langle x| \otimes |B\rangle \langle B|$, and
$S_{2}=\sum_{x}|x-1\rangle \langle x| \otimes |A\rangle \langle A|+ |x\rangle \langle x| \otimes |B\rangle \langle B|$.
In the presence of boundaries, quantum walks governed by $U$ exhibit non-Hermitian skin effects, as state evolutions are found to be localized near boundaries, due to the localization of all eigenstates.

In a discrete-time quantum walk, $U$ is repeatedly enforced upon an initial state, such that the time evolution constitutes a stroboscopic simulation of the dynamics governed by an effective Hamiltonian $H$, with $U=e^{-iH}$. Such a time evolution can be further regarded as a quench process if the initial state is an eigenstate of a certain $U^i$, or equivalently, an eigenstate of $H^i$ with  $U^i=e^{-iH^i}$. In Fig.~\ref{fig:fig4}(a), we show the topological phase diagram of the effective Hamiltonian under OBC. The phase diagram is the same as that of $U$ in the time frame given by Eq.~(\ref{eq:U}) and under OBC. Note that the GBZ of $U$ is always circular, which simplifies the problem.

To facilitate experimental implementation, a quasi-local initial state is preferred, which, to simultaneously fulfil the conditions of our scheme outlined above, should correspond to an eigenstate of $U^i$. As a concrete example, we choose a $U^i$ with $\theta^i_1=0.75\pi$ and $\theta^i_2=0.44\pi$ [point $I$ in Fig.~\ref{fig:fig4}(a)], and with a quasi-local eigenstate $|\Psi^i\rangle\propto |0\rangle\otimes(|A\rangle+|B\rangle)+2i/r_i|-1\rangle\otimes|A\rangle+1/r_i^2|-2\rangle\otimes(-|A\rangle+|B\rangle)$.
Writing the initial state in the GBZ as $|\Psi^i\rangle=\sum_k |\beta^i_{k,R}\rangle\otimes|\psi^i_k\rangle$ and evolve it repeatedly with $U^f$, we follow the prior practice of projecting the time-evolved density matrix into different $k$-sectors of the GBZ of $H^f$ (with $U^f=e^{-iH^f}$), from which $\bm{n}(k,t)$ can be extracted~\cite{supp}. We then vary parameters of $H^f$ along the line $I\rightarrow F$ in Fig.~\ref{fig:fig4}(a), and show the change of $\Delta \bar{n}_3=\bar{n}_3(k=0)-\bar{n}_3(k=\pi)$ in Fig.~\ref{fig:fig4}(b). Despite the deviation of $r_i/r_f$ from unity [see inset of Fig.~\ref{fig:fig4}(b)], $\Delta \bar{n}_3$ features an abrupt jump at the non-Bloch topological phase transition, thus providing a sensitive and robust signal for the detection of non-Bloch winding number $\nu_f$ of the effective Hamiltonian, once $\nu_i$ is known.

{\it Discussion:---}
We show that projecting quench dynamics onto the GBZ reveals dynamic topological structures that can be used as signals
for detecting non-Bloch topological invariants.
We explicitly demonstrate how such a non-Bloch quench dynamics can be simulated using quantum walks, which are experimentally accessible.
In the spirit of the non-Bloch quench dynamic discussed here, it is hopeful that various previously observed dynamic topological structures in unitary quench processes, such as vortices~\cite{Weitenberg17}, links~\cite{Weitenberg18}, and rings~\cite{chenshuai}, can also be explored in systems with non-Hermitian skin effects, which would then serve to detect non-Bloch topological invariants for more general models in higher dimensions.

\begin{acknowledgements} We acknowledge helpful discussions with Tian-Shu Deng. This work is supported by the National Natural Science Foundation of China (Grant Nos. 11974331, 11674306, 92065113) and the National Key R\&D Program (Grant Nos. 2016YFA0301700, 2017YFA0304100).
\end{acknowledgements}


\clearpage
\begin{widetext}
\appendix

\renewcommand{\thesection}{\Alph{section}}
\renewcommand{\thefigure}{S\arabic{figure}}
\renewcommand{\thetable}{S\Roman{table}}
\setcounter{figure}{0}
\renewcommand{\theequation}{S\arabic{equation}}
\setcounter{equation}{0}

\section{Supplemental Materials}

Here we provide more details on the non-Hermitian Su-Schieffer-Heeger (SSH) model, the projection of quench dynamics in the generalized Brillouin zone (GBZ), derivation for the conditional existence of fixed points and dynamic Chern numbers, analysis and numerical calculations demonstrating the persistence of skyrmions-like structures in more general settings, as well as theoretical and numerical analysis of quantum-walk dynamics.

\section{Non-Hermitian SSH model with skin effects}

For the non-Hermitian SSH model [Eq.~(1) in the main text] under the open boundary condition (OBC), all of its nominal bulk eigenstates are localized near the boundaries (for $\gamma\neq 0$)~\cite{WZ1}, and Bloch quasi-momenta of the Brillouin zone (BZ) are no longer a good bulk quantum numbers even in the thermodynamic limit. Central to our analysis here is the projection of dynamics onto GBZs characterized by the generalized momentum $k$ associated with the biorthogonal spatial basis
\begin{align}
|\beta_{k,R(L)} \rangle=\frac{1}{\sqrt{N}}\sum_{n} r^{\pm n} e^{ikn}|n\rangle,\label{eq:Sbeta}
\end{align}
where $r$ is the radius the GBZ, $k$ is identified as the generalized momentum of the GBZ, and the basis states satisfy $\langle\beta_{k,L}|\beta_{k',R} \rangle=\delta_{k,k'}$ and $\sum_{k}|\beta_{k,R}\rangle \langle\beta_{k,L}|=1$. While any single-particle state of the system can be generically written as $|\Psi\rangle=\sum_k |\beta_{k,R}\rangle\otimes |\psi_k\rangle$, with $|\psi_k\rangle=\langle \beta_{k,L}|\Psi\rangle$ its internal state in the sublattice space $\{|A\rangle,|B\rangle\}$, we introduce the projection operators $P_k=|\beta_{k,R}\rangle\langle \beta_{k,L}|\otimes \cp{1}_\alpha$ and $ Q_k=\cp 1- P_k$, where $\cp{1}_\alpha$ and $\cp 1$ are the identity operators in the sublattice space and the full Hilbert space of the system, respectively. This enables us to write, in the thermodynamic limit, $H=P_kHP_k+Q_kHQ_k=H_k+Q_kHQ_k$, where $H$ is given in Eq.~(1) of the main text, $H_k=P_kHP_k$, and we have used the fact that $P_kHQ_k$ and $Q_kHP_k$ vanish in the thermodynamic limit, though neither of them are exactly zero for a finite-size system.

Under the sublattice basis defined above, $H_k$ can be written in the matrix form
\begin{align}
H_k=
\begin{pmatrix}
0& t_{1}+\gamma+t_{2}r^{-1}e^{-ik}\\
  t_{1}-\gamma+t_{2}re^{ik} & 0
\end{pmatrix}.
\end{align}
The right and left eigenstates of $H_k$ are defined through $H_k|\psi_{k,\pm}\rangle=\pm E_{k}|\psi_{k,\pm}\rangle$ and $H^\dag_k|\chi_{k,\pm}\rangle=\pm E^\ast_{k}|\chi_{k,\pm}\rangle$, respectively. They further satisfy the biorthogonal and completeness relations:
$\langle \chi_{k,\mu}|\psi_{k,\nu}\rangle=\delta_{\mu\nu}$ and $\sum_{\mu} |\psi_{k,\mu}\rangle\langle\chi_{k,\mu}|=\cp{1}_{\alpha}$.
Throughout the work, we mainly focus on the regime with $t_1>\gamma$, where, under an exact non-Bloch parity-time (PT) symmetry, the eigen spectrum $E_k$ is completely real. Here $E_{k}= \sqrt{\overline{t}_{1}^2+t_{2}^2+2\overline{t}_{1}t_2\cos k}$, with $\overline{t}_{1}=\sqrt{(t_{1}-\gamma)(t_{1}+\gamma)}$, and $\pm$ are the band indices. The eigenstates are given as
\begin{align}
|\psi_{k,\pm}\rangle&=
\begin{pmatrix}
\pm 1/\sqrt{\overline{t}_{1}^2+t_{2}^2+2\overline{t}_{1}t_{2}\cos k}\\
  r/(\overline{t}_{1}+t_{2}e^{-ik})
\end{pmatrix},\\
|\chi_{k,\pm}\rangle&=
\begin{pmatrix}
\pm1/\sqrt{\overline{t}_{1}^2+t_{2}^2+2\overline{t}_{1}t_{2}\cos k}\\ 1/r(\overline{t}_{1}+t_{2}e^{-ik})
\end{pmatrix}.
\end{align}

While the biorthogonal spatial basis of the GBZ can be seen as the non-Bloch analog of Bloch waves, the complication of non-Bloch quench dynamics, i.e., the analysis of quench dynamics in the GBZ, comes from the general non-orthogonality of spatial basis states of the pre- and post-quench Hamiltonians, even if they are of the same $k$-sector.

\section{Quench dynamics and fixed points in the GBZ}

Consider a quench process initialized in the lower band of the initial Hamiltonian $H^i$, i.e., with all eigenstates $|\psi^i_{k,-}\rangle$ occupied. The initial state in a given $k$-sector is then $|\Psi^i_k\rangle=|\beta^i_{k,R}\rangle\otimes |\psi^i_{k,-}\rangle$, where $|\beta^i_{k,R}\rangle$ is the right spatial basis characterized, according to Eq.~(\ref{eq:Sbeta}), by the GBZ radius $r_i$ of $H^i$.
Evolving the initial state with the Hamiltonian $H^f$, the time-dependent density matrix is given by
\begin{align}
\rho(t)=e^{-iH^f t}|\Psi^i_k\rangle\langle\Psi^i_k| e^{iH^{f\dag} t}.
\end{align}
Importantly, when $\gamma\neq 0$, $\langle \beta^f_{k,L}|\beta^i_{k',R}\rangle\neq 0$ (and $\langle \beta^f_{k,R}|\beta^i_{k',L}\rangle\neq 0$), and dynamics in different $k$-sectors are coupled. Here $|\beta^{f}_{k,R/L}\rangle$ are the right/left spatial bases associated with the GBZ of $H^f$ according to Eq.~(\ref{eq:Sbeta}).

To overcome this difficulty, we project the dynamics onto the GBZ of $H^f$, such that the density matrix in each $k$-sector becomes
\begin{align}
\rho(k,t)=&P_k\rho(t)P_k^\dagger\nonumber\\
=&{\mathcal{N}^2}e^{-i H^f_kt}  |\psi^i_{k,-}\rangle\langle \psi^i_{k,-}| e^{i H^{f\dagger}_kt}
\end{align}
where $P_k=|\beta^f_{k,R}\rangle\langle \beta^f_{k,L}|\otimes \cp{1}_\alpha$, $Q_k=\cp 1-P_k$, $\mathcal{N}=\langle \beta^f_{k,L}|\beta^i_{k,R}\rangle=\frac{1}{N}\sum_{n} (\frac{r_{i}}{r_f})^n$, and we have used $H^f_k=P_k H^f P_k$, $P_k H^f Q_k=0$.

As discussed in the main text, the projected dynamics can be visualized on a Bloch sphere through
\begin{align}
\bm{n}(k,t)=\frac{\text{Tr}[\rho(k,t)\eta\bm{\tau}]}{\text{Tr}[\rho(k,t)\eta]},\label{eq:Snvec}
\end{align}
where $\eta=\sum_{\mu}|\chi^f_{k,\mu}\rangle\langle \chi^f_{k,\mu}|$ is the metric operator, $\bm{\tau} =(\tau_1, \tau_2, \tau_3)$, with $\tau_i=\sum_{\mu\nu=\pm}|\psi^f_{k,\mu}\rangle \sigma^{\mu\nu}_i \langle \chi^f_{k,\nu}|$ (i=0, 1, 2, 3). These definitions ensure that $\bm{n}(k,t)=(n_1,n_2,n_3)$ is always a real, unit vector.

When the eigen spectra $E_{k}^f$ of $H^f$ is real, we have~\cite{XPNC,iS}
\begin{align}
n_1=&(c_+^*c_-e^{2iE_k^ft}+c.c.)/n_0,\\
n_2=&i(c_+^*c_-e^{2iE_k^ft}-c.c.)/n_0,\\
n_3=&(c_+^*c_+-c_-^*c_-)/n_0,
\end{align}
where $c_{\mu}(k)=\langle \chi^f_{k,\mu}|\psi^i_{k,-}\rangle$, and $n_0=c_+^*c_++c_-^*c_-$. In this case, $n_3$ is time independent, and $\bm{n}(k,t)$ in a given $k$-sector rotates around the north pole of the Bloch sphere with a period $T_k=\pi/E^f_k$ [see Fig.~1(a) of the main text], where the north and south poles are given by the fixed-point conditions $c_-(k_m)=0$ and $c_+(k_m)=0$, respectively. Here $k_m$ is the location of fixed point, where local density matrix does not evolve in time.

When $E_k^f$ is purely imaginary,
\begin{align}
n_1=&(c_+^*c_-+c.c.)/n_0,\\
n_2=&i(c_+^*c_--c.c.)/n_0,\\
n_3=&(c_+^*c_+e^{-i2E_k^ft}-c_-^*c_-e^{i2E_k^ft})/n_0,
\end{align}
where $n_0=c_+^*c_+e^{-i2E_k^ft}+c_-^*c_-e^{i2E_k^ft}$. As $t$ tends to infinity, $\bm{n}(k,t)$ asymptotically approaches either the north (for $\text{Im} E_k>0$) or south (for $\text{Im} E_k<0$) pole [see Fig.~1(b) of the main text].

Further, from the analytical expression of $c_{\mu}$
\begin{align}
c_{\pm}(k)=\mp \frac{1}{\sqrt{\overline{t}_{1i}^2+t_{2i}^2+2\overline{t}_{1i}t_{2i}\cos k}}\frac{1}{\sqrt{\overline{t}_{1f}^2+t_{2f}^2+2\overline{t}_{1f}t_{2f}\cos k}}
+\frac{r_i}{r_f}\frac{1}{(\overline{t}_{1i}+t_{2i}e^{-ik})(\overline{t}_{1f}+t_{2f}e^{ik})},\label{eq:Sc}
\end{align}
we see that when $r_i=r_f$, there are always two fixed points at $k_m=0$ and $k_m=\pi$, respectively.

The existence of fixed points and periodic dynamics in the $k$-sectors (under a completely real $E^f_k$) divide the generalized momentum-time space into a series of $S^2$ submanifolds, on each of which a quantized, dynamic Chern number can be defined
\begin{align}
C_{mn}=\frac{1}{4\pi} \int^{k_n}_{k_m}dk\int^{T_k}_{0}dt   [\bm{n} (k,t) \times \partial _t \bm{n}(k,t) ] \cdot \partial _k \bm{n} (k,t),
\end{align}
where $k_m,k_n\in\{0,\pm\pi\}$ are two adjacent fixed points. It is straightforward to show that: when $c_+(k_m)=0$ and $c_-(k_n)=0$, $C_{mn}=1$; when $c_-(k_m)=0$ and $c_+(k_n)=0$, $C_{mn}=-1$; otherwise, $C_{mn}=0$. Hence, if we categorize fixed points into two different kinds, with either $c_+(k_m)=0$ or $c_-(k_m)=0$, the dynamic Chern number can only take non-zero values when the corresponding submanifold is pinned by fixed points of different kinds. While these dynamic Chern numbers serve as skyrmion numbers for the dynamic skyrmions in the generalized momentum-time space shown in the main text, they are linked with the non-Bloch winding numbers of the pre- and post-quench Hamiltonians. Specifically, from Eq.~(\ref{eq:Sc}) and in combination with the expression for non-Bloch winding numbers [Eq.~(2) of the main text], we find that,
for $\nu_i\neq \nu_f$, two fixed points of different kinds emerge at $0,\pi$, ensuring the appearance of dynamic skyrmion structure. However, for $\nu_i=\nu_f$, the two fixed points at $0$ and $\pi$ are of the same kind, and dynamic skyrmions are absent. We note that Eq.~(\ref{eq:Sc}) as well sa the conclusions above are conditional on the reality of the eigen spectra $E^{i,f}_k$ of $H^{i,f}_k$, which is the regime of study here.

For the more general case of $r_i\neq r_f$, i.e., the GBZs of the pre- and post-quench Hamiltonians do not match, fixed points do not exist in general. Firstly, this can be concluded by observing that $c_\pm(k)\neq 0$ in Eq.~(\ref{eq:Sc}). Furthermore, as a state initialized in a given $k$-sector would proliferate into other $k$-sectors during the time evolution, Eq.~(\ref{eq:Sc}) cannot be used to characterize dynamics for the quench dynamics of a topological system at half-filling, where all $k$-sectors of the lower band are occupied initially at $t=0$.
However, as long as $r_i/r_f$ does not significantly deviate from unity, the impact of the these cross coupling terms are typically small [see for instance, a related analysis following Eq.~(\ref{eq:SHeff})]; and in each $k$-sector, $\bm{n}(0,t)$ and $\bm{n}(\pi,t)$ (for completely real $E^f_k$) precess closely around the poles of the Bloch sphere, and fixed points exist in a perturbative sense. Although dynamic Chern numbers cannot be rigorously defined in this case, global signatures of dynamic skyrmions (for $\nu_i\neq\nu_f$), or the lack there of (for $\nu_i=\nu_f$), would still be discernable beyond $r_i=r_f$. In fact, in the following section, we show that dynamic skyrmions even persist when GBZs of neither Hamiltonians are circular. Such an understanding underlies our proposal for a general detection scheme of non-Bloch topological invariants.

\section{Non-Hermitian SSH model with next-nearest neighbor hopping }

To further illustrate the applicability of our scheme on models with more general GBZs, we consider the quench dynamics of a non-Hermitian SSH model with next-nearest neighbor hopping, i.e., with an additional term
\begin{align}
H_{nn}=\sum_n (t_3|n,A\rangle\langle n+1,B|+H.c.),
\end{align}
where $t_3$ is the hopping rate. For a non-Hermitian SSH model with a finite $t_3$, the GBZ is no longer a circle on the complex plane. The spatial basis involves a $k$-dependent radius
\begin{align}
|\beta_{k,R(L)} \rangle=\frac{1}{\sqrt{N}}\sum_{n} r^{\pm n}(k) e^{ikn}|n\rangle, \quad k \in(-\pi,\pi],
\end{align}
which does not satisfy the biothorgonal relations in the main text, i.e., $\langle\beta_{k',L}|\beta_{k,R} \rangle\neq 0$. Further, non-Bloch PT symmetry is broken with a finite $t_3$, such that the eigen spectrum under OBC is always complex, leading to steady-state approaching quench dynamics in general.

However, when $t_3$ is small, the GBZ does not deviate too much from a circle, remnants of the dynamic skymion structures still exist for $\nu_i\neq\nu_f$, and can still be used to determine the post-quench non-Bloch winding number once that of the initial Hamiltonian is known.

More concretely, the projected Hamiltonian in a given $k$-sector is given by
\begin{align}
H_k=
\begin{pmatrix}
0& t_{1}+\gamma+t_{2}r^{-1}e^{-ik}+t_{3}re^{ik}\\
  t_{1}-\gamma+t_{2}re^{ik}+t_{3}r^{-1}e^{-ik} & 0
\end{pmatrix},
\end{align}
and one can perform the same analysis outlined above to study projected quench dynamics in the GBZ of $H^f$.

In Fig.~\ref{fig:figS1}, we show the numerical simulation of such a quench process, with the state in each $k$-sector of the GBZ initialized in
the lower-band of $H^i_k$, and projecting the dynamics onto the GBZ of $H^f$. In Fig.~\ref{fig:figS1}(a), $\nu_i=0$ and $\nu_f=1$, so that skyrmion-like structures are visible. By contrast, in Fig.~\ref{fig:figS1}(b), $\nu_i=1$ and $\nu_f=1$, and no visible skyrmion structures are present. Furthermore, we characterize in Fig.~\ref{fig:figS1}(c) the variation of GBZs with increasing $t_3$, and show the resulting $\bar{n}_3$ as a function of $t_3$ in Fig.~\ref{fig:figS1}(d). Similar to the main text, $\bar{n}_3=(1/T^\prime_k)\int_0^{T^\prime_k}[n_3(0,t)-n_3(\pi,t)]dt$, where $T^\prime_k=\text{Re}E_k^f$, is used as the signature for the skyrmion-like structure. Apparently, $\bar{n}_3\sim 2$ ($\bar{n}_3\sim 0$) for $\nu_i\neq\nu_f$ ($\nu_i=\nu_f$), for over a considerable range of $t_3$.

\begin{figure}[htbp]
\centering
\includegraphics[width=6.5in]{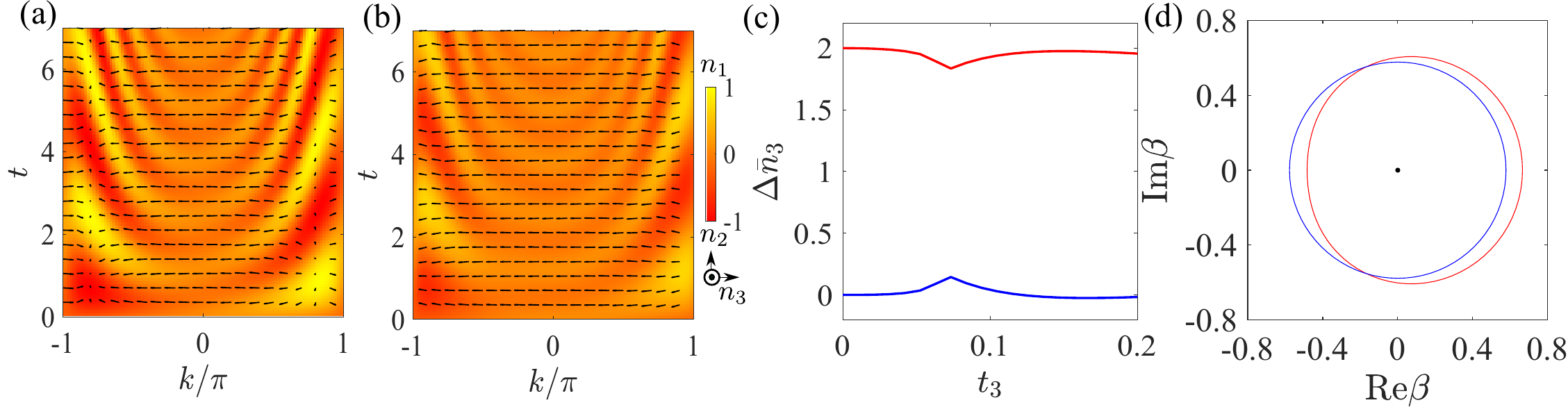}
\caption{(a)(b) Spin texture $\bm{n}(k,t)$ on the generalized momentum-time space for non-Bloch quench dynamics under non-Hermitian SSH model with next-nearest-neighbor hopping, with $\gamma=0.5$. (a) Quench from $H^i$ with $t^i_2=0.5$ and $t^i_3=0$ (non-Bloch winding number $\nu^i=0$) to $H^f$ with $t^f_2=1.5$ and $t^f_3=0.05$ (non-Bloch winding number $\nu^f=1$). (b) Quench from $t^i_2=1$ and $t^i_3=0$ (non-Bloch winding number $\nu^i=1$) to $t^f_2=1.5$ and $t^f_3=0.05$.
(c) $\Delta\bar{n}_3$ with increasing $t^f_3$ and a fixed $t^f_2=1.5$. The initial parameters are $t^i_2=0.5$ (red) and $t^i_2=1$ (blue), respectively. (d) GBZs of $H^f$ in (c), with $t^f_3=0$ (blue) and $t^f_3=0.05$ (red), respectively.
GBZs of $H^i$ for both cases coincide with the blue circle. We take $t_1=1$ as the unit of energy.
}
\label{fig:figS1}
\end{figure}

\section{Quench dynamics in a quantum walk}
\begin{figure*}[tbp]
\centering
\includegraphics[width=7in]{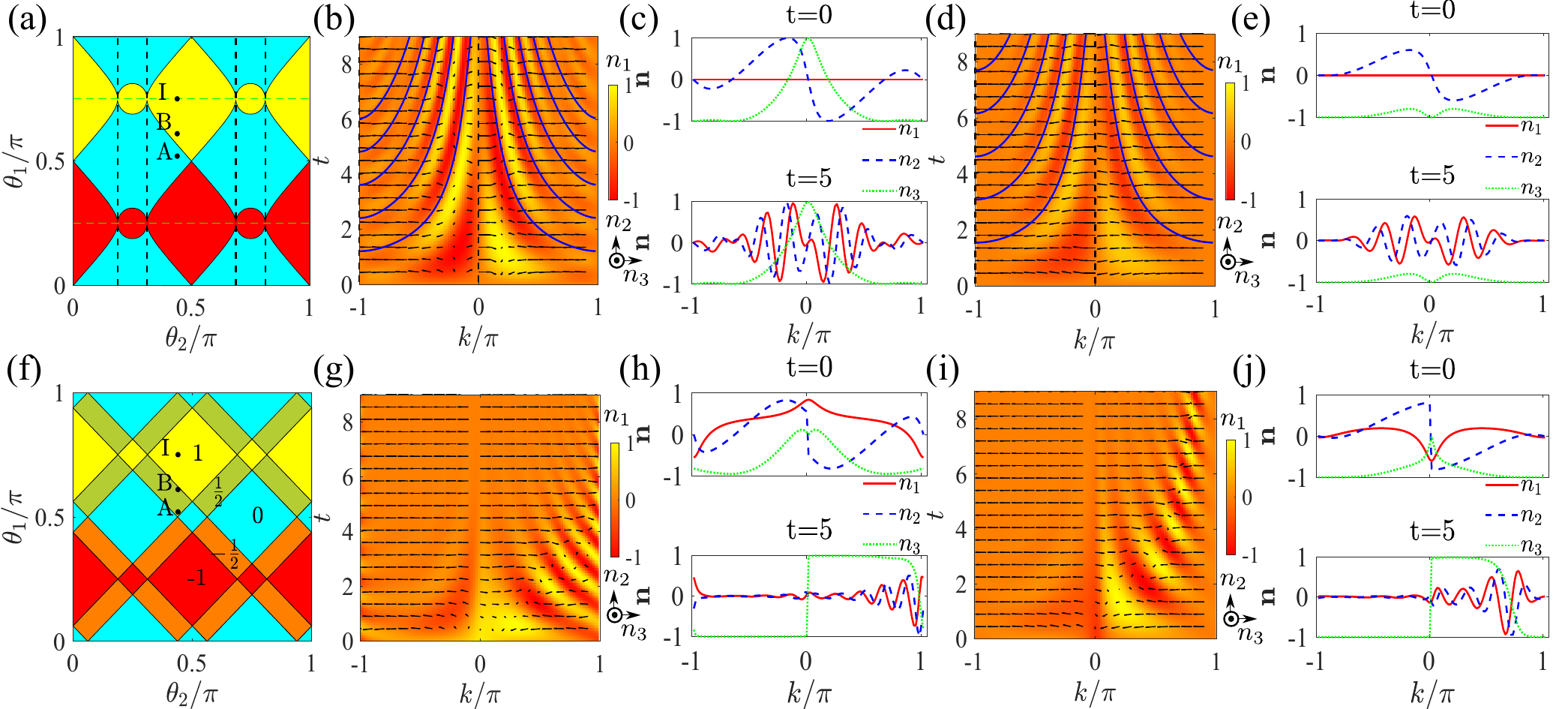}
\caption{
Stroboscopic simulation of quench dynamics using a quantum walk governed by $U$ in Eq.~(6), with $\gamma=0.4$. Quench dynamics is projected in (top) the GBZ and (lower) BZ, respectively. (a) Non-Bloch topological phase diagram of the effective Hamiltonian $H$ under OBC, where $U=e^{-iH}$.
The horizontal green lines represent locations of localized eigenstates. The vertical dashed lines are the non-Bloch exceptional lines separating the exact non-Bloch PT symmetric and PT-broken regimes, where we only focus on the PT-unbroken regime. Regions filled in yellow, cyan and red respectively feature non-Bloch winding numbers $\nu=1,0,-1$.
(b) Dynamic spin texture $\bm{n}(k,t)$ in the generalized momentum-time space, for a quantum quench from point I (with non-Bloch winding number $\nu_i=1$) to A (with non-Bloch winding number $\nu_f=0$). (c) $\bm{n}(k,t)$ in the GBZ  at different times $t$ for the quench in (b). (d) Spin texture $\bm{n}(k,t)$ for a quench from point I to B (with non-Bloch winding number $\nu_f=1$). No skyrmion structures are present, in contrast to (b). (e) Components of $\bm{n}(k,t)$ in the GBZ at different times $t$ for the quench in (d). (f) Bloch topological phase diagram under PBC. Eigen spectra are complex throughout the phase diagram. Bloch winding numbers are marked in the phase diagram, where regions with different winding numbers are colored differently. (g) Dynamic spin texture $\bm{n}(k,t)$ in the momentum-time space (BZ), for a quantum quench from point I (with Bloch winding number $\nu_i=1$) to A (with Bloch winding number $\nu_f=1/2$). (h) Components of $\bm{n}(k,t)$ in the BZ at different times $t$ for the quench in (g).
(i) Spin texture $\bm{n}(k,t)$ in the momentum-time space for a quench from point I to B (with Bloch winding number $\nu_f=1/2$). (j) Components of $\bm{n}(k,t)$ in the BZ at different times $t$ for the quench in (i). The parameters for points $A$, $B$, and $I$ are $(\theta_1,\theta_2)=(0.52\pi,0.44\pi), (0.61\pi, 0.44\pi), (0.75\pi, 0.44\pi)$, respectively.
}
\label{fig:figS2}
\end{figure*}

As discussed in the main text, we consider quantum walk dynamics governed by the Floquet operator $U$ given by Eq.~(6). Under an OBC, the GBZ of $U$ is a circle on the complex plane, its radius $r=\sqrt{|\frac{\cosh \gamma \cos 2\theta_{2}-\sinh \gamma}{\cosh \gamma \cos 2\theta_{2}+\sinh \gamma}|}$~\cite{EP}.

We start from a localized initial state $|\Psi^i\rangle\propto|0,A\rangle+|0,B\rangle+2i/r_i|-1,A\rangle+1/r_i^2(-|-2,A\rangle+|-2,B\rangle)$, which is an eigenstate of an initial Floquet operator $U^i$ with $\theta^i_1=3\pi/4$ and $\theta^i_2=0.44\pi$. Formally writing the initial state as
$|\Psi_i\rangle=1/\sqrt N\sum_{k} |\Psi^i_k\rangle$, where $|\Psi^i_k\rangle=|\beta^i_{k,R}\rangle\otimes|\psi^i_k\rangle$,
we evolve it under the Floquet operator $U=e^{-iH^f t}$. The time-evolved state is then $|\Psi(t)\rangle=\sum_k e^{-iH^ft}|\Psi^i_k\rangle$. Projecting the dynamics onto the GBZ of the final effective Hamiltonian $H^f$, we have
\begin{align}
|\psi(k,t)\rangle &= P_k|\Psi(t)\rangle= P_k \sum_k e^{-iH^{f}t}|\Psi^i_k\rangle \nonumber\\
&= \mathcal{N}(0)(e^{-iH^f_kt}|\psi^i_k\rangle+\sum_{k'\neq k}\frac{\mathcal{N}(k-k')}{\mathcal{N}(0)} e^{-iH^f_kt}|\psi^i_{k'}\rangle),
\label{eq:SHeff}
\end{align}
where $ \mathcal{N}(k-k')=\langle \beta^f_{k,L}|\beta^i_{k',R}\rangle=\frac{1}{N}\sum_{n} (\frac{r_{i}}{r_f})^n e^{-i(k-k')n} $.
Since $|\mathcal{N}(k-k')/\mathcal{N}(0)|\sim |(r_i-r_f)/(r_i+r_f)|$, the second term on the right-hand side of Eq.~(\ref{eq:SHeff}) vanishes for $r_i=r_f$, and is negligibly small for $|r_i-r_f|\ll r_i+r_f$. We note that a similar situation should occur for the quench dynamics of lattice models at half filling, as we discussed previously.

To see the relation between non-Bloch winding numbers of the effective Hamiltonian and non-Bloch quench dynamics, we first project the Floquet operator $U$ onto the GBZ
\begin{align}
U(k)=P_kUP_k,
\end{align}
which, under the parameters considered in the main text $|\cos2\theta_2|>|\tanh\gamma|$, can be decomposed as
\begin{align}
U(k)=d_0 \sigma_0 -id_1\sigma_1-id_2\sigma_2-id_3\sigma_3,
\end{align}
where
\begin{align}
&d_0=\lambda\cos2\theta_1  \cos k-\sin2\theta_1\sin2\theta_2\cosh\gamma,\\
&d_1=0,\\
&d_2=\lambda\sin2\theta_1  \cos k-\cos2\theta_1 \sin2\theta_2\cosh\gamma,\\
&d_3=-\lambda \sin k.,
\end{align}
with $\lambda=[(\cosh \gamma \cos 2\theta_{2}-\sinh \gamma)(\cosh \gamma \cos 2\theta_{2}+\sinh \gamma)]^{1/2}$.

The non-Bloch winding number of the effective Hamiltonian H, with $U=e^{-iH}$, is calculated through~\cite{photonskin}
\begin{align}
\nu=\frac{1}{2\pi}\int dk \frac{-d_3\partial_k d_2+d_2\partial_k d_3}{d_2^2+d_3^2},\label{eq:Snonblochnu}
\end{align}
which reduces to the Bloch winding number for $\gamma=0$.

Notably, $U(k)$ is unitary and chiral symmetric with $\sigma_x U(k)\sigma_x=U^{-1}(k)$, where $\sigma_x$ is the chiral symmetry operator. It follows that $H^f_k$ in Eq.~(\ref{eq:SHeff}) is Hermitian and chiral symmetric, and the proof regarding the existence of topological fixed points and dynamic Chern number (hence dynamic skyrmions) is essentially the same as that in Ref.~\cite{Chen17} . Importantly, these conclusions are exact for $r_i=r_f$, when the second term in Eq.~(\ref{eq:SHeff}) is zero, and approximately true when the term is negligibly small.

In Fig.~\ref{fig:figS2}, we perform a numerical simulation of quantum walk dynamics, starting from a localized state in the bulk.
The quench dynamics is projected onto the GBZ in the first row of Fig.~\ref{fig:figS2}, and onto the BZ in the second row. We focus here on the case with $r_i=r_f$, as more general cases are already shown in the main text. Here dynamic skyrmions emerge in Fig.~\ref{fig:figS2}(b), where $\nu_i=0$ and $\nu_f=1$. In Fig.~\ref{fig:figS2}(d), on the other hand, no dynamic skyrmions are present, as $\nu_i=1$ and $\nu_f=1$. Therefore, if either the non-Bloch winding number of the initial or the final effective Hamiltonian is known, one can infer the value of the other one by the observation of spin textures. Consistent with these conclusions, we find that fixed points exist at $k_m=0$ and $k_m=\pi$ for both cases [see Fig.~\ref{fig:figS2}(c)(e)], but they are of the same kind in Fig.~\ref{fig:figS2}(d)(e), hence the dynamic Chern numbers are zero therein.

By contrast, when projected on the BZ, as shown in Fig.~\ref{fig:figS2}(f-j), the spin textures feature no skyrmion structures, and the dynamics is steady-state approaching in all $k$-sectors.


\end{widetext}

\end{document}